\documentclass{nature}

\usepackage{lettrine}
\usepackage{graphicx}
\usepackage{dcolumn}
\usepackage{bm}
\usepackage{amssymb}
\usepackage{amssymb}
\usepackage{epsfig}
\usepackage[10pt]{moresize}
\usepackage{comment}
\def\>{\rangle}

\title{Quantum internet using code division multiple access}

\author{Jing Zhang$^{1,2,3\star}$, Yu-xi Liu$^{1,3,4}$, \c{S}ahin Kaya \"{O}zdemir$^{1,5}$, Re-Bing Wu$^{1,2,3}$, Feifei Gao$^{2,3}$, Xiang-Bin Wang$^{1,6}$, Lan Yang$^{5}$ \& Franco Nori$^{1,7}$}

\begin{document}

\maketitle

\begin{affiliations}
 \item CEMS, RIKEN, Saitama, 351-0198, Japan
 \item Department of Automation, Tsinghua
University, Beijing 100084, P. R. China
  \item Center for Quantum
Information Science and Technology, TNList, Beijing 100084, P. R.
China
  \item Institute of Microelectronics, Tsinghua University,
Beijing 100084, P. R. China
  \item Electrical and Systems
Engineering, Washington University, St. Louis, Missouri 63130,
USA
  \item Department of Physics, Tsinghua University, Beijing
100084, P. R. China
  \item Physics Department, The University of
Michigan, Ann Arbor, MI 48109-1040,
USA
\\$^\star$e-mail:jing-zhang@mail.tsinghua.edu.cn
\end{affiliations}

\begin{abstract}
A crucial open problem in large-scale quantum networks is how to
efficiently transmit quantum data among many pairs of users via a
common data-transmission medium. We propose a solution by
developing a quantum code division multiple access (q-CDMA)
approach in which quantum information is chaotically encoded to
spread its spectral content, and then decoded via chaos
synchronization to separate different sender-receiver pairs. In
comparison to other existing approaches, such as frequency
division multiple access (FDMA), the proposed q-CDMA can greatly
increase the information rates per channel used, especially for
very noisy quantum channels.
\end{abstract}

\lettrine[lines=2]{Q}uantum networks for long distance communication and distributed
computing require nodes which are capable of storing and
processing quantum information and connected to each other via
photonic channels.\cite{Kimble,Felinto,Chou}. Recent achievements
in quantum
information\cite{Lukin,Sangouard,You,You2,Clark,Buluta,Blatt} have
brought quantum networking much closer to realization. Quantum
networks exhibit advantages when transmitting classical and
quantum information with proper encoding into and decoding from
quantum states\cite{Yard,Horodecki,Ralph,Meter,Zbinden,Wang,Pan}.
However, the efficient transfer of quantum information among many
nodes has remained as a problem yet to be
solved\cite{Mabuchi,Maitre,Phillips,Duan,Matsukevich,Acin,Lu},
which becomes more crucial for the limited-resource scenarios in
large-scale networks. Multiple access, which allows simultaneous
transmission of multiple quantum data streams in a shared channel,
may provide a remedy to this problem.

Popular multiple-access methods in classical communication
networks include time-division multiple-access (TDMA),
frequency-division multiple-access (FDMA), and code-division
multiple-access (CDMA). See Fig.~1 for an illustration of
different multiple-access methods. In TDMA, different users share
the same frequency but transmit on different time slots, but
timing synchronization and delays become serious problems in
large-scale networks. In FDMA, different users share the same time
slots but operate on different frequency bands. However, only a
narrow band of the data transmission line has a low leakage rate
and the bands assigned to different users should be sufficiently
separated to suppress interference. Unlike TDMA and FDMA, CDMA
utilizes the entire spectrum and time slots to encode the
information for all users, while distinguishes different users
with their own unique codes. Therefore, CDMA is adopted as the key
technology of the currently-used third generation mobile
communication systems, and can accommodate more bits per channel
use\cite{TMCover} compared with TDMA and FDMA. It has achieved
great success in commercial applications of classical
communications.

Although FDMA has already been used in quantum key distribution
networks\cite{Fujiwara,Bussieres,Ortigosa-Blanch,Godbout,Townsend},
to the best of our knowledge, CDMA has not yet been applied in
quantum networks and internet\cite{Kimble}. A q-CDMA network would
require that the states sent by each transmitting node of the
quantum network are encoded into their coherent superposition
before being sent to the common channel, and the quantum
information for each of the intended receiving node is coherently
and faithfully extracted by proper decoding at the end of the
common channel. This, however, is not a trivial task but rather a
difficult one.

In this paper, we propose a q-CDMA method via chaotic encoding and
chaos synchronization among senders and receivers, which require a
quantum channel to transmit quantum superposition states and N
classical channels for chaos synchronization to decode the quantum
signals at the receiver nodes. It can be seen that the proposed
q-CDMA provides higher transmission rates for both classical and
quantum information, especially in very noisy channels.

\section*{Results}

To present the underlying principle of our method, we consider the
simplest case, where two pairs of sender and receiver nodes
communicate quantum information, encoded into quantized
electromagnetic fields with the {\it same} frequencies, via a
single quantum channel [see Fig.~2(a)].

The schematic diagram of our strategy is shown in Fig.~2(b). The
quantum information sent by the nodes $1$ and $2$ is first encoded
by two chaotic phase shifters ${\rm CPS}_1$ and ${\rm CPS}_2$,
whose operation can be modelled by the effective Hamiltonian
$\delta_i\!\left(t\right)a^{\dagger}_i a_i$, with
$\delta_i\!\left(t\right)$ being time dependent classical chaotic
signals and $i=1,2$. This encoding spreads the spectral content of
the quantum information across the entire spectrum. The two beams
are then combined at the $50\!:\!50$ beamsplitter ${\rm BS}_1$ and
transmitted via a common channel to the receivers. At the end of
the channel, the quantum signal is first amplified by a
phase-insensitive linear amplifier (LA), then divided into two
branches by a second $50\!:\!50$ beamsplitter ${\rm BS}_2$, and
finally sent to nodes $3$ and $4$ through two chaotic phase
shifters ${\rm CPS}_3$ and ${\rm CPS}_4$, which are introduced to
decode the information by applying the effective Hamiltonian
$-\delta_j\!\left(t\right)a_j^{\dagger}a_j$, with $j=3,4$.
Amplifier gain is set as $G=4$ to compensate the losses induced by
the beamsplitters.

The actions of the chaotic devices ${\rm CPS}_{i=1,2,3,4}$ induce
the phase shifts $\exp\left[-i\theta_i\!\left(t\right)\right]$,
where $\theta_i\!\left(t\right)=\int_0^t\delta_i\left(\tau\right)d
\tau$. Thus, to achieve faithful transmission between the senders
and the receivers, the effects of $\delta_1\!\left(t\right)$ and
$\delta_2\!\left(t\right)$ on the quantum signals should be
minimized in the fields received by the nodes $3$ and $4$.
Intuitively, this could be done by simply adjusting the system
parameters such that
$\delta_1\!\left(t\right)=\delta_3\!\left(t\right)$ and
$\delta_2\!\left(t\right)=\delta_4\!\left(t\right)$. However, such
an approach is impractical, because any small deviation in the
system parameters can be greatly amplified by the chaotic motion,
making it impossible to keep two chaotic circuits with the same
exact parameters and initial conditions. Instead, auxiliary
classical channels between senders and the intended receivers can
be used to synchronize the chaotic circuit as shown in Fig.~2(b).
This classical chaotic synchronization helps to reduce the
parameter differences between the chaotic phase shifters and to
extract the quantum information faithfully.

\subsection{Modelling of quantum CDMA network.} Hereafter, for the
sake of simplicity, we assume that ${\rm CPS}_1$ (${\rm CPS}_2$)
and ${\rm CPS}_3$ (${\rm CPS}_4$) have been synchronized before
the start of the transmission of quantum information, i.e.,
$\theta_1\!\left(t\right)=\theta_3\!\left(t\right)$
[$\theta_2\!\left(t\right)=\theta_4\!\left(t\right)$]. The whole
information transmission process in this quantum network can be
described by the input-output relationship
\begin{eqnarray}\label{Input-output relationship between the senders and receivers}
a_3&=&a_1+a_2
e^{i\left(\theta_1-\theta_2\right)}+\frac{\sqrt{6}}{2}e^{i\theta_1}a_{\rm
LA}^{\dagger}+\frac{1}{\sqrt{2}}e^{i\theta_1}a_{\rm
BS},\nonumber\\
a_4&=&a_2+a_1
e^{i\left(\theta_2-\theta_1\right)}+\frac{\sqrt{6}}{2}e^{i\theta_2}a_{\rm
LA}^{\dagger}-\frac{1}{\sqrt{2}}e^{i\theta_2}a_{\rm BS},
\end{eqnarray}
where $a_{\rm LA}^{\dagger}$ and $a_{\rm BS}$ are the creation and
annihilation operators of the auxiliary vacuum fields entering the
linear amplifier LA and the second beamsplitter ${\rm BS}_2$. For
the pseudo-noise chaotic phase-shift $\theta_i\left(t\right)$, we
should take an average over this broadband random signal, which
leads to $\overline{\exp\left(\pm
i\theta_i\left(t\right)\right)}\approx\sqrt{M_i}$ with
\begin{equation}\label{Chaotic correction factor}
M_i=\exp\left[-\pi\int_{\omega_{li}}^{\omega_{ui}}\!\!d\omega\;
S_{\delta_i}\!\left(\omega\right)/\omega^2\right].
\end{equation}
In Eq.~(\ref{Chaotic correction factor}),
$S_{\delta_i}\!\left(\omega\right)$ is the power spectrum density
of the signal $\delta_i\!\left(t\right)$, and $\omega_{li}$ and
$\omega_{ui}$ are the lower and upper bounds of the frequency band
of $\delta_i\!\left(t\right)$, respectively. Equation
(\ref{Input-output relationship between the senders and
receivers}) can then be reduced to
\begin{eqnarray}\label{Input-output relationship between the senders and receiver: reduction}
a_3&=&a_1+\sqrt{M_1 M_2}\:a_2+\sqrt{\frac{3 M_1}{2}}a_{\rm
LA}^{\dagger}+\sqrt{\frac{M_1}{2}}a_{\rm
BS},\nonumber\\
a_4&=&a_2+\sqrt{M_1 M_2}\:a_1+\sqrt{\frac{3 M_1}{2}}a_{\rm
LA}^{\dagger}-\sqrt{\frac{M_2}{2}}a_{\rm BS}.
\end{eqnarray}
For a chaotic signal with broadband frequency spectrum, the factor
$M_i$ is extremely small, and can be neglected in
Eq.~(\ref{Input-output relationship between the senders and
receiver: reduction}). This leads to $a_3 \approx a_1$ and $a_4
\approx a_2$, implying efficient and faithful transmission of
quantum information from nodes $1$ and $2$ to nodes $3$ and $4$,
respectively.

In our q-CDMA network, the information-bearing fields $a_1$ and
$a_2$, having the same frequency $\omega_c$, are modulated by two
different pseudo-noise signals, which not only broaden them in the
frequency domain but also change the shape of their wavepackets
[see Fig.~2(b)]. Thus, the energies of the fields $a_1$ and $a_2$
are distributed over a very broad frequency span, in which the
contribution of $\omega_c$ is extremely small and impossible to
extract without coherent sharpening of the $\omega_c$ components.
This, on the other hand, is possible only via {\it chaos}
synchronization which effectively eliminates the pseudo-noises in
the fields and enables the recovery of $a_1$ ($a_2$) at the output
$a_3$ ($a_4$) with almost no disturbance from $a_2$ ($a_1$). This
is similar to the classical CDMA. Thus, we name our protocol as
q-CDMA.

\subsection{Quantum state transmission.} Let us further study
the transmission of qubit states over the proposed q-CDMA network
using a concrete model. The qubit states
\{$|\phi_1\rangle=\sqrt{p_1} | g_1 \rangle + \sqrt{1-p_1} | e_1
\rangle$, and $| \phi_2 \rangle = \sqrt{p_2} | g_2 \rangle +
\sqrt{1-p_2} | e_2 \rangle$, with $p_1,\,p_2 \in
\left[0,1\right]$\}, to be transmitted are encoded in the dark
states of two $\Lambda$-type three-level atoms; i.e., atom $1$ in
cavity $1$ and atom $2$, in cavity $2$, as shown in Fig.~3(a). The
qubit states are transferred to the cavities by Raman transitions
and are transmitted over the q-CDMA network. At the receiver
nodes, the quantum states are transferred and stored in two
$\Lambda$-type atoms; i.e., atom $3$ in cavity $3$, and atom $4$
in cavity $4$. We assume that the four coupled atom-cavity systems
have the same parameters. Let $|g_i\rangle$, $|e_i\rangle$, and $|
r_i \rangle$ be the three energy levels of atom $i$. As shown in
Fig.~3(a), the $|g_i\rangle \rightarrow | r_i \rangle$ and $| e_i
\rangle \rightarrow | r_i \rangle$ transitions are coupled with a
classical control field and a quantized cavity field with coupling
strengths $\Omega_i \left( t \right)$ and $g_i \left( t \right)$.
By adiabatically eliminating the highest energy level $| r_i
\rangle$, the Hamiltonian of the atom-cavity system can be
expressed as
\begin{equation}\label{Atom-cavity interaction Hamiltonian}
H_i=g_i\left(t\right)\left(c_i|e_i\rangle\langle
g_i|+c_i^{\dagger}|g_i\rangle\langle e_i|\right),
\end{equation}
where $c_i$ is the annihilation operator of the $i$-th cavity
mode; $g_i \left( t \right) = g \Omega_i \left( t \right) /
\Delta$ is the coupling strength which can be tuned by the
classical control field $\Omega_i \left( t \right)$; and $\Delta$
is the atom-cavity detuning. The cavity fields $c_i$ are related
to the travelling fields $a_i$ by
\begin{eqnarray}\label{Input-output relationship of the four cavities}
&a_1=\sqrt{\kappa}\;c_1+a_{1,{\rm in}}\,, \quad
a_2=\sqrt{\kappa}\;c_2+a_{2,{\rm in}}\,, & \nonumber\\
&a_{3,{\rm out}}=\sqrt{\kappa}\;c_3+a_3, \quad a_{4,{\rm
out}}=\sqrt{\kappa}\;c_4 + a_4, &
\end{eqnarray}
where $\kappa$ is the decay rate of the cavity field; and
$a_{1,{\rm in}}$, $a_{2,{\rm in}}$ (both in vacuum states) and
$a_{3,{\rm out}}$, $a_{4,{\rm out}}$ are the input and output
fields of the whole system, respectively.

The chaotic phase shifters ${\rm CPS}_{i=1,2,3,4}$ are realized by
coupling the optical fields to four driven Duffing oscillators,
with damping rates $\gamma$, described by the Hamiltonian
\begin{equation}\label{Hamiltonian of the Duffing oscillator}
H_{{\rm
Duff},i}=\frac{\omega_o}{2}p_i^2+\frac{\omega_o}{2}x_i^2-\mu
x_i^4-f\left(t\right) x_i,
\end{equation}
where $x_i$ and $p_i$ are the normalized position and momentum of
the nonlinear Duffing oscillators, $\omega_0/2\pi$ is the
frequency of the fundamental mode, $\mu$ is a nonlinear constant,
and  $f\!\left(t\right)=f_d\cos\left(\omega_d t\right)$ is the
driving force. The interaction between the field $a_i$ and the
$i$-th Duffing oscillator is given by the Hamiltonian
\begin{equation}\label{Optomechanical Hamiltonian}
H_i=g_{\rm f-o} x_i a_i^{\dagger} a_i,
\end{equation}
where $g_{\rm f-o}$ is the coupling strength between the field and
the oscillator. Under the semiclassical approximation for the
degrees of freedom of the oscillator, the interaction Hamiltonian
$H_i$ leads to a phase factor $\exp\left[-i\int_0^t g_{\rm  f-o}
x_i\!\left(\tau\right) d\tau \right]$ for the field $a_i$. To
simplify the discussion, we assume that all of the four Duffing
oscillators have the same $\omega_0$, $\mu$, $f_d$, and
$\omega_d$, but different initial states. Finally, the chaotic
synchronization between ${\rm CPS}_1$ (${\rm CPS}_2$) and ${\rm
CPS}_3$ (${\rm CPS}_4$) is achieved by coupling two Duffing
oscillators by a harmonic potential
$V\left(x_1,x_3\right)=k_I\left(x_1-x_3\right)^2/2$.

The nonlinear coupling between the optical fields and the Duffing
oscillators and the chaos synchronization to achieve the chaotic
encoding and decoding could be realized using different physical
platforms. For example, in optomechanical systems, the interaction
Hamiltonian~(\ref{Optomechanical Hamiltonian}) can be realized by
coupling the optical field via the radiation pressure to a moving
mirror connected to a nonlinear spring (see Fig. 3(b)). Chaotic
mechanical resonators can provide a frequency-spreading of several
hundreds of MHz for a quantum signal, and this is broad enough,
compared to the final recovered quantum signal, to realize the
q-CDMA and noise suppression. Chaos synchronization between
different nonlinear mechanical oscillators can be realized by
coupling the two oscillators via a linear spring. This kind of
synchronization of mechanical oscillators have been realized in
experiments\cite{Shim}, but it is not suitable or practical for
long-distance quantum communication. Chaos synchronization with a
mediating optical field, similar to that used to synchronize
chaotic semiconductor lasers for high speed secure
communication\cite{JOhtsubo}, would be the method of choice for
long-distance quantum communication. The main difficulty in this
method, however, will be the coupling between the classical
chaotic light and the information-bearing quantum light. This, on
the other hand, can be achieved via Kerr interactions. There is a
recent report\cite{Sahin} that proposes to use Kerr nonlinearity
in whispering gallery mode resonators to solve this problem.
Another approach for chaotic encoding and chaos synchronization
between distant nodes of the network could be the use of
electro-optic modulators (EOMs). See, e.g., Fig.~3(c). In this
case, the input information-bearing quantum signal is modulated by
the EOM driven by a chaotic electrical signal\cite{MTsang}. The
EOM can prepare the needed broadband signal, and there have been
various proven techniques of chaotic signal generation and
synchronization in electrical circuits. Indeed, recently
experimental demonstration of chaos synchronization in a four-node
optoelectronic network was reported\cite{CRSWilliams}.

To show the efficiency of state transmission in q-CDMA, let us
calculate the fidelities $F_1=\langle\phi_1|\rho_3|\phi_1\rangle$
and $F_2=\langle\phi_2|\rho_3|\phi_2\rangle$, where $\rho_3$ and
$\rho_4$ are the quantum states received by atoms $3$ and $4$, and
$|\phi_1 \rangle = \sqrt{ p_0 } | g_1 \rangle + \sqrt{ 1-p_0 } |
e_1 \rangle$ and $ | \phi_2 \rangle = \sqrt{ 1 - p_0 } | g_2
\rangle + \sqrt{ p_0 } | e_2 \rangle $ are the two quantum states
to be transmitted. By designing the control parameters
$g_i\left(t\right)$, using the Raman transition
technique\cite{Mabuchi}, we find for the particular chosen quantum
states that the fidelities $F_1$ and $F_2$ can be approximated as
$F_1 = F_2 \approx 1 - M$. When the Duffing oscillator enters the
chaotic regime, we have $ M \approx 0$, leading to fidelities
$F_1,\,F_2\approx 1$, which means that the qubit states can be
faithfully transmitted over the q-CDMA network.

We show the feasibility of the q-CDMA method using numerical
simulations with the system parameters $\omega_d/\omega_0=5$,
$g_{\rm f-o}/\omega_0=0.03$, $\mu/\omega_0=0.25$,
$\gamma/\omega_0=0.05$, $k_I/\omega_0=0.1$, and $p_0=0.6$. In
Fig.~4(a), it is seen that there are three distinct regions
representing how the chaotic motion affects the fidelity of the
quantum state transmission. In the periodic regime characterized
by $0<f_d/\omega_0<15$, both $F_1$ and $F_2$ experience slight
increases with increasing $f_d/\omega_0$, with $0.4<F_1<0.5$ and
$0.6\leq F_2\leq0.64$. At $f_d/\omega_0=15$, the Duffing
oscillator enters the soft chaotic regime which is indicated by a
positive Lyapunov exponential $\lambda \approx 0.038$ and a sudden
jump in fidelities. In this regime, delineated by $15 \leq
f_d/\omega_0 \leq 33$, both $F_1$ and $F_2$ are still below $0.7$.
The dynamics of the Duffing oscillator enters the hard-chaos
regime at $f_d/\omega_0\approx 33$, where both $F_1$ and $F_2$
suddenly jump to $1$, which corresponds to an almost $100\%$
faithful state transmission. In Fig.~4(b), we plot the
trajectories of $F_1$ and $F_2$ as a function of $p_0$ in the
hard-chaotic regime $f_d/\omega_0=36$, corresponding to $M\approx
0.0103$. It is seen that $F_1$ and $F_2$ are very close to $1-M
\approx0.9897$ and almost constant regardless of the value of
$p_0$. There are small deviations from $1-M$, because here $M^2$
terms are not neglected. The average fidelity
$\bar{F}=\left(F_1+F_2\right)/2$ is maximum at $p_0=1/2$, which
corresponds to an equally-weighted superposition of the quantum
states $|\phi_1\rangle$ and $|\phi_2\rangle$. In such a case, the
crosstalk between the channels becomes minimum, inducing only a
very slight disturbance on these indistinguishable states.

\subsection{Information transmission rates.} Next we consider
the maximum transmission rates of classical and quantum
information over the proposed q-CDMA network, and compare them,
under certain energy constraints, with the achievable bounds of
transmission rates in a q-FDMA network and in quantum networks
without any multiple access method (i.e., single user-pair
network). Here the classical information transmission rates are
calculated in terms of the Holevo information\cite{ASHolevo,Keyl}
and the quantum information transmission rates are defined by the
coherent information\cite{IDevetak,Lloyd,Shor}. We assume that the
frequencies allocated to different user pairs in the FDMA network
are equally spaced such that the number of users is maximized and
cross-talks between adjacent channels are suppressed. Moreover, we
restrict our discussion to {\it Gaussian channels} and {\it
Bosonic channels}, respectively for the transmissions of quantum
and classical information.

We briefly summarize the main results here and in Fig.~5(a)-(c).
(i) For lossless channels (i.e., $\eta=1$ where $\eta$ denotes the
transmissivity of the central frequency of the information-bearing
field), upper bounds of classical and the quantum information
transmission rates for the proposed q-CDMA network are higher than
those of the quantum FDMA and the single user-pair networks if the
crosstalk in the q-CDMA is suppressed by setting $M\ll 1$. (ii)
With the increasing number $N$ of user-pairs in the networks,
q-CDMA increasingly performs better than the q-FDMA for classical
and quantum information. (iii) Information transmission rates for
the q-CDMA is more robust to noise. For fixed $N$, quantum
information transmission rates of the q-FDMA and the single
user-pair networks degrades very fast to zero as the loss $1-\eta$
increases from zero (ideal channel) to $1/2$, whereas the q-CDMA
network retains its non-zero rate even for very noisy channels.
For the classical information transmission, the situation is
similar except that the transmission rates of q-FDMA and the
single user-pair network drops to zero when $\eta=0$ which
corresponds to a completely lossy channel.

The robustness of the proposed q-CDMA network for noisy channel
can be explained as follows. The chaotic phase shifters in the
q-CDMA network spread the information-bearing field across a broad
spectral band. Thus, the energy distributed in a particular mode
is almost negligible, and thus the photon loss is also almost
negligible. Therefore, increasing $\eta$ has very small effect on
the transmission rates. In Fig.~5(b)-(c), we consider the noise to
be broadband, and shows that the transmission rates of classical
and quantum information over the q-CDMA network change only
slightly.

\section*{Discussion}

We have introduced a q-CDMA network based on chaotic
synchronization where quantum information can be faithfully
transmitted with fidelities as high as $0.99$ between multiple
pairs of nodes sharing a single quantum channel. The proposed
quantum multiple-access network is robust against channel noises,
and attains higher transmission rates for both classical and
quantum information when compared to other approaches. A q-CDMA
network based on our proposal requires the realization of two
important issues. First, quantum interference of signals from
different chaotic sources. This has recently been demonstrated by
Nevet \emph{et. al.}~\cite{Nevet}. Second is the implementation of
chaotic phase shifters and their synchronization. These could be
implemented in various systems, including but not limited to
optomechanical, optoelectrical~\cite{MTsang}, and all-optical
systems~\cite{JOhtsubo}. In particular, whispering-gallery-mode
(WGM) optical resonators are possible platforms as chaotic
behavior in a WGM microtoroid resonator has been reported in
Ref.~\cite{TCarmon}. Although synchronization of self-sustaining
oscillations in directly coupled microring resonators have been
demonstrated~\cite{Zhang}, and mechanical mode synchronization in
two distant resonators coupled via waveguides has been
proposed~\cite{Manipatruni}, demonstration of chaos
synchronization in such optomechanical resonators are yet to be
demonstrated. Although the tasks to be fulfilled  are not trivial,
we believe that we are not far away from such realizations due to
the rapid pace of experimental and theoretical developments we
have seen in the field in the past few years. We think that our
proposal will pave the way for long distance q-CDMA networks, and
will give new perspectives for the optimization of quantum
networks.

\begin{methods}

\subsection{Averaging over the chaotic phase shift.}

A chaotic signal $\delta_i\!\left(t\right)$ can be expressed as a
combination of many high-frequency components, i.e.,
\begin{equation}
\delta_i\!\left(t\right)=\sum_{\alpha}A_{i\alpha}\cos\left(\omega_{i\alpha}t+\phi_{i\alpha}\right),
\end{equation}
where $A_{i\alpha}$, $\omega_{i\alpha}$, $\phi_{i\alpha}$ are the
amplitude, frequency, and phase of each component, respectively.
Then the phase of the signal at any given time $t$ can be written
as
\begin{eqnarray*}
\theta_i\!\left(t\right)=\int_0^t \delta_i\!\left( \tau \right) d
\tau =
\sum_{\alpha}\frac{A_{i\alpha}}{\omega_{i\alpha}}\sin\left(\omega_{i\alpha}t+\phi_{i\alpha}\right).
\end{eqnarray*}
Using the Fourier-Bessel series identity\cite{JZhang}:
\begin{eqnarray*}
\exp{\left(ix\sin y\right)}=\sum_n
J_n\!\left(x\right)\exp{\left(iny\right)},
\end{eqnarray*}
with $J_n\!\left(x\right)$ as the $n$-th Bessel function of the
first kind, we can write
\begin{eqnarray*}
\exp{\left[-i\theta_i\!\left(t\right)\right]}=\prod_{\alpha}\left[\sum_{n_{\alpha}}J_{n_{\alpha}}\!\!\left(\frac{A_{i\alpha}}{\omega_{i\alpha}}\right)e^{-in_{\alpha}\omega_{i\alpha}t-i
n_{\alpha}\phi_{i\alpha}}\right].
\end{eqnarray*}
If we take average over the ``random" phase
$\theta_i\left(t\right)$, the components related to the
frequencies $\omega_{i\alpha}$ should appear as fast-oscillating
terms and thus can be averaged out. This treatment corresponds to
averaging out the components that are far off-resonance with the
information-bearing field, and keeping only the near-resonance
components. Hence, only the lowest-frequency terms, with
$n_{\alpha}=0$, dominate the dynamical evolution. Thus, we have
\begin{equation}\label{Averaging over the chaotic phase}
\overline{\exp{\left[-i\theta_i\!\left(t\right)\right]}}=\prod_{\alpha}\left[J_0\!\left(\frac{A_{i\alpha}}{\omega_{i\alpha}}\right)\right].
\end{equation}
Since the chaotic signal $\delta_i\!\left(t\right)$ is mainly
distributed in the high-frequency regime, we have
$A_{i\alpha}\ll\omega_{i\alpha}$. Using the expressions
$J_0\!\left(x\right)\approx1-x^2/4$, $\log\left(1+x\right) \approx
x$ for $x \ll 1$, it is easy to show that
\begin{eqnarray}\label{Calculation of the correction factor}
\prod_{\alpha}J_0\!\left(\frac{A_{i\alpha}}{\omega_{i\alpha}}\right)&=&\exp\left[\sum_{\alpha}\log
J_0\!\left(\frac{A_{i\alpha}}{\omega_{i\alpha}}\right)\right]\nonumber\\
&=&\exp\left(-\frac{1}{4}\sum_{\alpha}\frac{A_{i\alpha}^2}{\omega_{i\alpha}^2}\right)\nonumber\\
&=&\exp\left(-\frac{\pi}{2}\int_{\omega_{li}}^{\omega_{ui}}\frac{S_{\delta_i}\left(\omega\right)}{\omega^2}d\omega\right)\nonumber\\
&=&\sqrt{M_i},
\end{eqnarray}
where
\begin{eqnarray*}
M_i=\exp\left(-\pi\int_{\omega_{li}}^{\omega_{ui}}\frac{S_{\delta_i}\left(\omega\right)}{\omega^2}d\omega\right).
\end{eqnarray*}
Consequently, from Eqs.~(\ref{Averaging over the chaotic phase})
and (\ref{Calculation of the correction factor}), we obtain the
equation
\begin{equation}\label{Averaging over the chaotic phase to
obtain the correction factor}
\overline{\exp{\left[-i\theta_i\left(t\right)\right]}}=\sqrt{M_i}.
\end{equation}

\subsection{Input-output relationship of the quantum CDMA
network.}

Here we calculate the input-output relationship of the quantum
CDMA network shown in Fig.~6, we can express the input-output
relationships of the chaotic phase shifters ${\rm
CPS}_{i=1,2,3,4}$  as
\begin{eqnarray}\label{Input-output relationship of CPS}
& a_1'=a_1 e^{-i \theta_1},\quad a_2'=a_2 e^{-i \theta_2}, &
\nonumber\\
& a_3=a_3' e^{i \theta_1}, \quad a_4=a_4' e^{i \theta_2}, &
\end{eqnarray}
and those of the two beam splitters ${\rm BS}_1$ and ${\rm BS}_2$
and the linear quantum amplifier ``LA", respectively, as

\begin{equation}\label{Input-output relationship of BS1}
a_5=\frac{1}{\sqrt{2}}a_1'+\frac{1}{\sqrt{2}}a_2',\quad
a_6=\frac{1}{\sqrt{2}}a_1'-\frac{1}{\sqrt{2}}a_2',
\end{equation}
\begin{equation}\label{Input-output relationship of LA}
a_7=2 a_5 + \sqrt{3} a_{\rm LA}^{\dagger}, \quad a_8=\sqrt{3}
a_5^{\dagger}+2 a_{\rm LA},
\end{equation}
and
\begin{equation}\label{Input-output relationship of BS2}
a_3'=\frac{1}{\sqrt{2}}a_7+\frac{1}{\sqrt{2}}a_{\rm BS},\quad
a_4'=\frac{1}{\sqrt{2}}a_7-\frac{1}{\sqrt{2}}a_{\rm BS}.
\end{equation}
Then, using Eqs.~(\ref{Input-output relationship of
CPS}-\ref{Input-output relationship of BS2}), we obtain the total
input-output relationship of the quantum network as
\begin{eqnarray}\label{Input-output relationship of the
quantum network with the chaotic phase shifters} a_3&=&a_1+a_2
e^{i\left(\theta_1-\theta_2\right)}+\frac{\sqrt{6}}{2}e^{i\theta_1}a_{\rm
LA}^{\dagger}+\frac{1}{\sqrt{2}}e^{i\theta_1}a_{\rm
BS},\nonumber\\
a_4&=&a_2+a_1
e^{i\left(\theta_2-\theta_1\right)}+\frac{\sqrt{6}}{2}e^{i\theta_2}a_{\rm
LA}^{\dagger}-\frac{1}{\sqrt{2}}e^{i\theta_2}a_{\rm BS}.
\end{eqnarray}
where $\theta_1$ and $\theta_2$ are independent chaotic ``noises"
as we have not considered chaos synchronization yet.

\end{methods}


\begin{thebibliography}{1}

\bibitem{Kimble} Kimble, H.~J. The quantum internet. {\it Nature} {\bf 453}, 1023-1030 (2008).

\bibitem{Felinto} Felinto D. et al. Conditional control of the
quantum states of remote atomic memories for quantum networking.
{\it Nature Phys.} {\bf 2}, 844-848 (2006).

\bibitem{Chou} Chou, C.-W. et al. Functional quantum nodes for entanglement distribution over scalable quantum networks. {\it Science} {\bf 316}, 1316-1320 (2007).

\bibitem{Lukin} Lukin, M. D. Colloquium: Trapping and manipulating photon states in atomic
ensembles. {\it Rev. Mod. Phys.} {\bf 75}, 457-472 (2003)

\bibitem{Sangouard} Sangouard, N., Simon, C., Riedmatten, H. de \& Gisin, N. Quantum repeaters based on atomic ensembles and linear optics. {\it Rev.
Mod. Phys.} {\bf 83}, 33-80 (2011).

\bibitem{You} You, J.~Q. \& Nori, F. Superconducting circuits and quantum information. {\it Physics Today} {\bf 58} (11),
42-47 (2005).

\bibitem{You2} You, J.~Q. \& Nori F. Atomic physics and quantum optics using superconducting
circuits. {\it Nature} {\bf 474}, 589-597 (2011).

\bibitem{Clark} Clarke, J. \& Wilhelm, F.~K. Quantum bits. {\it Nature} {\bf 453}, 1031-1042 (2008).

\bibitem{Buluta} Buluta, I., Ashhab, S. \& Nori, F. Natural and artificial atoms for quantum computation. {\it Rep. Prog. Phys.} {\bf 74},
104401 (2011).

\bibitem{Blatt} Blatt R. \& Wineland, D.~J. Entangled states of trapped atomic ions. {\it Nature} {\bf 453}, 1008-1014 (2008).

\bibitem{Yard} Smith G. \& Yard, J. Quantum communication with zero-capacity channels. {\it Science} {\bf 321}, 1812-1815
(2008).

\bibitem{Horodecki} Czekaj, L. \& Horodecki, P. Purely quantum superadditivity of classical capacities of quantum multiple access channels. {\it Phys. Rev. Lett.} {\bf 102}, 110505
(2009).

\bibitem{Ralph} Heurs M. et al. Multiplexed communication over a high-speed quantum channel. {\it Phys. Rev. A} {\bf 81}, 032325 (2010).

\bibitem{Meter} Aparicio L. \& Meter, R. V. Multiplexing schemes for quantum repeater networks. {\it Proc. SPIE} {\bf 8163}, 816308
(2011).

\bibitem{Zbinden} Gisin, N., Ribordy, G., Tittel, W. \& Zbinden, H. Quantum cryptography. {\it Rev. Mod. Phys.} {\bf 74}, 145-195 (2002).

\bibitem{Wang} Wang, X.~B., Hiroshima, T., Tomita, A. \& Hayashi, M. Quantum information with Gaussian states. {\it Phys.
Rep.} {\bf 448}, 1-111 (2007).

\bibitem{Pan} Pan, J.-W. et al. Multiphoton entanglement and interferometry. {\it
Rev. Mod. Phys.} {\bf 84}, 777-838 (2012).

\bibitem{Mabuchi} Cirac, J.~I., Zoller, P., Kimble, H.~J. \& Mabuchi, H. Quantum state transfer and entanglement distribution among distant nodes in a quantum network. {\it Phys. Rev. Lett.} {\bf 78},
3221-3224 (1997).

\bibitem{Maitre} Maitre, X. et al. Quantum memory with a single photon in a cavity. {\it Phys. Rev. Lett.} {\bf 79}, 769-772 (1997).

\bibitem{Phillips} Phillips, D.~F., Fleischhauer, A., Mair, A., Walsworth, R. L. \& Lukin, M. D. Storage of light in atomic vapor. {\it Phys. Rev.
Lett.} {\bf 86}, 783-786 (2001).

\bibitem{Duan} Duan, L.~M., Lukin, M.~D., Cirac, J.~I. \& Zoller, P. Long-distance quantum communication with atomic ensembles and linear optics. {\it
Nature} {\bf 414}, 413-418 (2001).

\bibitem{Matsukevich} Matsukevich D.~N. \&  Kuzmich, A. Quantum state transfer between matter and
light. {\it Science} {\bf 306}, 663-666 (2004).

\bibitem{Acin} Ac\'{i}n, A., Cirac, J.~I. \& Lewenstein, M. Entanglement percolation in quantum networks. {\it Nature Phys.} {\bf
3}, 256-259 (2007).

\bibitem{Lu} L\"{u}, X.~Y., Liu, J.~B., Ding, C.~L. \& Li, J.~H. Dispersive atom-field interaction scheme for three-dimensional entanglement between two spatially separated atoms. {\it Phys. Rev.
A}
{\bf 78}, 032305 (2008).

\bibitem{TMCover} Cover T.~M. \& Thomas, J.~A. {\it Elements of Information Theory} (Wiley,  New York, 1991), page 407.

\bibitem{Fujiwara} Yoshino, K. et al. High-speed wavelength-division multiplexing quantum key
distribution system. {\it Opt. Lett.} {\bf 37}, 223-225 (2012).

\bibitem{Bussieres} Brassard, G., Bussieres, F., Godbout, N. \& Lacroix, S. Entanglement and wavelength division multiplexing for quantum cryptography networks. {\it AIP
Conf. Proc.} {\bf 734}, 323-326 (2004).

\bibitem{Ortigosa-Blanch} Ortigosa-Blanch, A. \& Capmany, J. Subcarrier multiplexing optical quantum key distribution. {\it Phys. Rev. A} {\bf 73},
024305 (2006).

\bibitem{Godbout} Brassard, G., Bussieres, F., Godbout, N. \& Lacroix, S. Multiuser quantum key distribution using
wavelength division multiplexing. {\it Proc. SPIE} {\bf 5260},
149-153 (2003).

\bibitem{Townsend} Townsend, P. Simultaneous quantum cryptographic key distribution and conventional data transmission over installed fibre using wavelength-division multiplexing. {\it Electron. Lett.} {\bf 33}, 188-190 (1997).

\bibitem{JZhang} Zhang, J. et al. Deterministic
chaos can act as a decoherence suppressor. {\it Phys. Rev. B} {\bf
84}, 214304 (2011).

\bibitem{Shim} Shim, S.-B., Imboden, M. \& Mohanty, P. Synchronized oscillation in coupled nanomechanical oscillators. {\it Science} {\bf 316}, 95-99 (2007).

\bibitem{JOhtsubo} Ohtsubo, J. Chaos synchronization and chaotic signal masking in semiconductor lasers with optical feedback. {\it IEEE J. Quantum Electronics} {\bf 38}, 1141-1154 (2002).

\bibitem{Sahin} Xiao, Y.~F., \"{O}zdemir, S.~K., Gaddam, V., Dong, C.~H., Imoto, N. \& Yang, L. Quantum nondemolition measurement of photon number via optical Kerr effect in an ultra-high- microtoroid cavity Opt.
Exp. {\bf 16}, 21462 (2008).

\bibitem{MTsang} Tsang, M. Cavity quantum electro-optics. II. Input-output relations between traveling optical and microwave fields. {\it Phys. Rev. A} {\bf 84}, 043845 (2011).

\bibitem{CRSWilliams} Williams, C.R.S. et al. Experimental observations of group synchrony in
a system of chaotic optoelectronic oscillators. {\it Phys. Rev.
Lett.} {\bf 110}, 064104 (2013).

\bibitem{ASHolevo} Holevo, A.~S. \& Werner, R.~F. Evaluating capacities of bosonic Gaussian channels. {\it Phys. Rev. A} {\bf
63}, 032312 (2001).

\bibitem{Keyl} Keyl, M. Fundamentals of Quantum Information Theory. {\it Phys. Rep.} {\bf 369}, 431-548 (2002).

\bibitem{IDevetak} Devetak, I. The private classical capacity and quantum capacity of a quantum channel. {\it IEEE Trans. Inf. Theory} {\bf 51},
44-55 (2005).

\bibitem{Lloyd} Lloyd, S. Capacity of the noisy quantum channel. Phys. Rev. A {\bf 55}, 1613-1622 (1997).

\bibitem{Shor} Shor, P.~W. The quantum channel capacity and coherent information. Lecture Notes, MSRI Workshop on Quantum Computation (2002).

\bibitem{Nevet} Nevet, A., Hayat, A., Ginzburg, P. \& Orenstein, M. Indistinguishable photon pairs from independent true chaotic sources. {\it Phys. Rev. Lett.} {\bf 107}, 253601
(2011).

\bibitem{TCarmon} Carmon, T., Cross, M.~C. \& Vahala, K.~J. Chaotic quivering of micron-scaled on-chip resonators excited by centrifugal optical pressure. {\it Phys. Rev. Lett.} {\bf
98}, 167203 (2007).

\bibitem{Zhang} Zhang, M. et al. Synchronization of micromechanical oscillators using light. {\it Phys. Rev. Lett.} {\bf 109}, 233906 (2012).

\bibitem{Manipatruni} Manipatruni, S., Wiederhecker, G. \& Lipson, M. Long-range synchronization of optomechanical structures. {\it Quantum electronics and laser
science Conference}, Baltimore, Maryland, May 2011.

\end{thebibliography}


\begin{addendum}
 \item[Acknowledgement] JZ thanks Dr. X.Y.L\"{u} and Prof M. Tsang for helpful discussions, and also thanks Dr. G. Y. Chen for providing related materials. All authors thank F. Monifi for the illustrations in Fig.~1. FN is partially supported by the ARO, RIKEN iTHES Project, NSF grant No.0726909, JSPS-RFBR contract No.12-02-92100, Grant-in-Aid for Scientific Research (S), MEXT Kakenhi on Quantum Cybernetics, and the JSPS via its FIRST program. JZ and RBW are supported by the NSFC under Grant Nos.61174084, 61134008, 60904034. YXL is supported by the NSFC under Grant Nos.~10975080, 61025022, 60836001. LY is supported by ARO grant No. W911NF-12-1-0026.
 \item[Correspondence] Correspondence and requests for materials should be addressed to
J.Z.
 \item[Author Contributions] J.Z. proposed the main idea. J.Z., S.K.O., Y.X.L., and F.N. wrote the main manuscript text. R.B.W., F.F.G, X.B.W., and L.Y. participated in discussing the results and contributed to the findings of this paper.
 \item[Competing Interests] The authors declare that they have no
competing financial interests.
\end{addendum}


\clearpage

\textbf{Figure 1: Illustration for different multiple-access
methods.} (a) TDMA: the users share the same frequency at
different time slots. (b) FDMA: different frequency bands are
assigned to different data-streams. (c) CDMA: the entire spectrum
is utilized to encode the information from all users, and
different users are distinguished with their own unique codes.
Each user in the network is represented by a different color.
\bigskip

\textbf{Figure 2: Diagrams of the quantum multiple access
networks.} (a) Quantum information transmission between two pairs
of nodes via a single quantum channel. Quantum states from two
senders are combined to form a superposition state and input to
the channel. At the receiver side, they are coherently split into
two and sent to the targeted receivers. (b) Schematic diagram of
the q-CDMA network by chaotic synchronization. Wave packets from
the sender nodes are first spectrally broadened by using the
chaotic phase shifters ${\rm CPS}_1$ and ${\rm CPS}_2$, and then
mixed at a beamsplitter (${\rm BS}_1$) and input to the channel.
After linear amplification (LA) and splitting at the second
beamsplitter (${\rm BS}_2$), individual signals are recovered at
the receiver end with the help of ${\rm CPS}_3$ and ${\rm CPS}_4$,
which are synchronized with ${\rm CPS}_1$ and ${\rm CPS}_2$,
respectively.
\bigskip

\textbf{Figure 3: Quantum state transmission over q-CDMA network.}
(a) The broom-shaped or shovel-shaped purple symbols denote photon
detectors. The red arrow inside each (green) cavity denotes the
classical driving field with amplitude $\Omega_i \left( t \right)$
($i=1,2,3,4$). The green circles denote $\Lambda$-type three-level
atoms. (b) Schematic diagram of the chaotic synchronization
realized by the moving mirrors. (c) Chaotic encoding and decoding
by electro-optic modulators.
\bigskip

\textbf{Figure 4: Fidelities of quantum state transmission.} (a)
Fidelities $F_1$ and $F_2$ versus the strength $f_d$ of the
driving force acting on the Duffing oscillator with $p_0=0.6$, and
$\tau=2\pi/\omega_0$ as the unit of time. (b) $F_1$, $F_2$ and
their average $\left(F_1+F_2\right)/2$ versus $p_0$ in the
hard-chaotic region, with $f_d/\omega_0=36$. The average fidelity
is maximized at $p_0=0.5$, which corresponds to
$|\phi_1\rangle=|\phi_2\rangle$.
\bigskip

\textbf{Figure 5: Quantum information transmission rates.} (a)
Fidelities $F_1$ and $F_2$ versus the strength $f_d$ of the
driving force acting on the Duffing oscillator with $p_0=0.6$, and
$\tau=2\pi/\omega_0$ as the unit of time. (b) Upper bounds of the
classical and quantum information transmission rates of different
methods for ideal channel with $\eta=1$ versus the number of the
user pairs $N$. (c) and (d) Upper bounds of classical (quantum)
information transmission rates of different methods for noisy
channel with $0<\eta<1$. The correction factor in the q-CDMA
network is $M=0.01$. FDMA is constrained by the frequency
bandwidth $\delta\omega/\omega=0.2$. All the methods are
constrained with the total energy $\epsilon/\omega=1$. $C_{\rm
c(q),CDMA(FDMA)}^{\eta}$ denote the classical (c) and quantum (q)
information transmission rates in q-CDMA and q-FDMA networks with
transmissivity $\eta$. The rates for the single user-pair channel
are $C_{\rm c,sig}^{\eta}$ and $C_{\rm q,sig}^{\eta}$.
\bigskip

\textbf{Figure 6: Input-output structure of quantum CDMA netwrok.}
The black dashed lines denote the desired chaotic synchronization
channel. The red lines show the quantum optical channels. ``LA"
refers to linear amplifier. ``BS" refers to beamsplitter. ``CPS"
denotes chaotic phase shifter.
\bigskip

\clearpage
\begin{figure}
\begin{center}
\epsfig{file=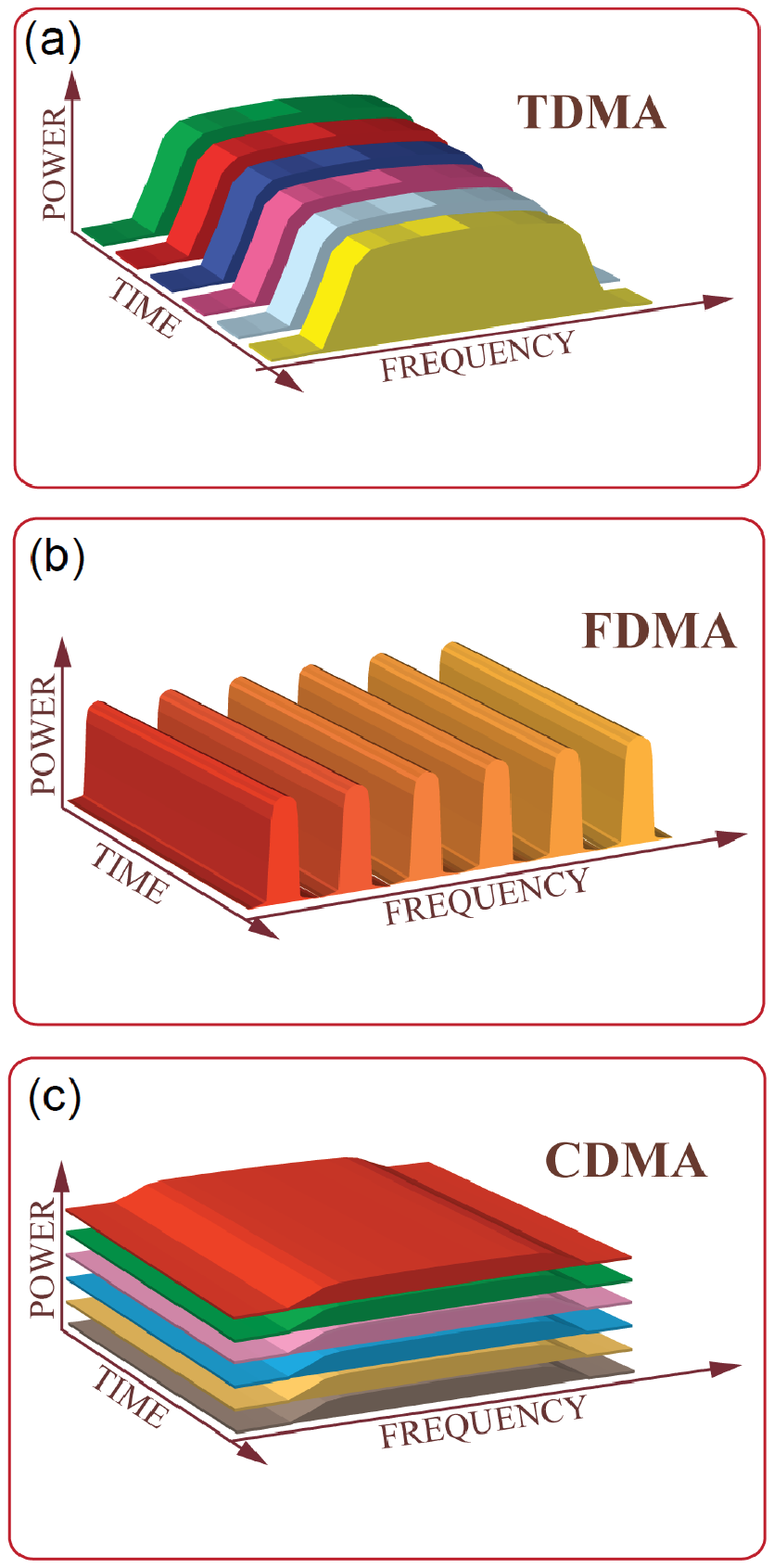,width=10cm}
\end{center}\caption{}
\label{Fig of the schematic diagram of quantum multiple access
network}
\end{figure}

\clearpage
\begin{figure}
\begin{center}
\epsfig{file=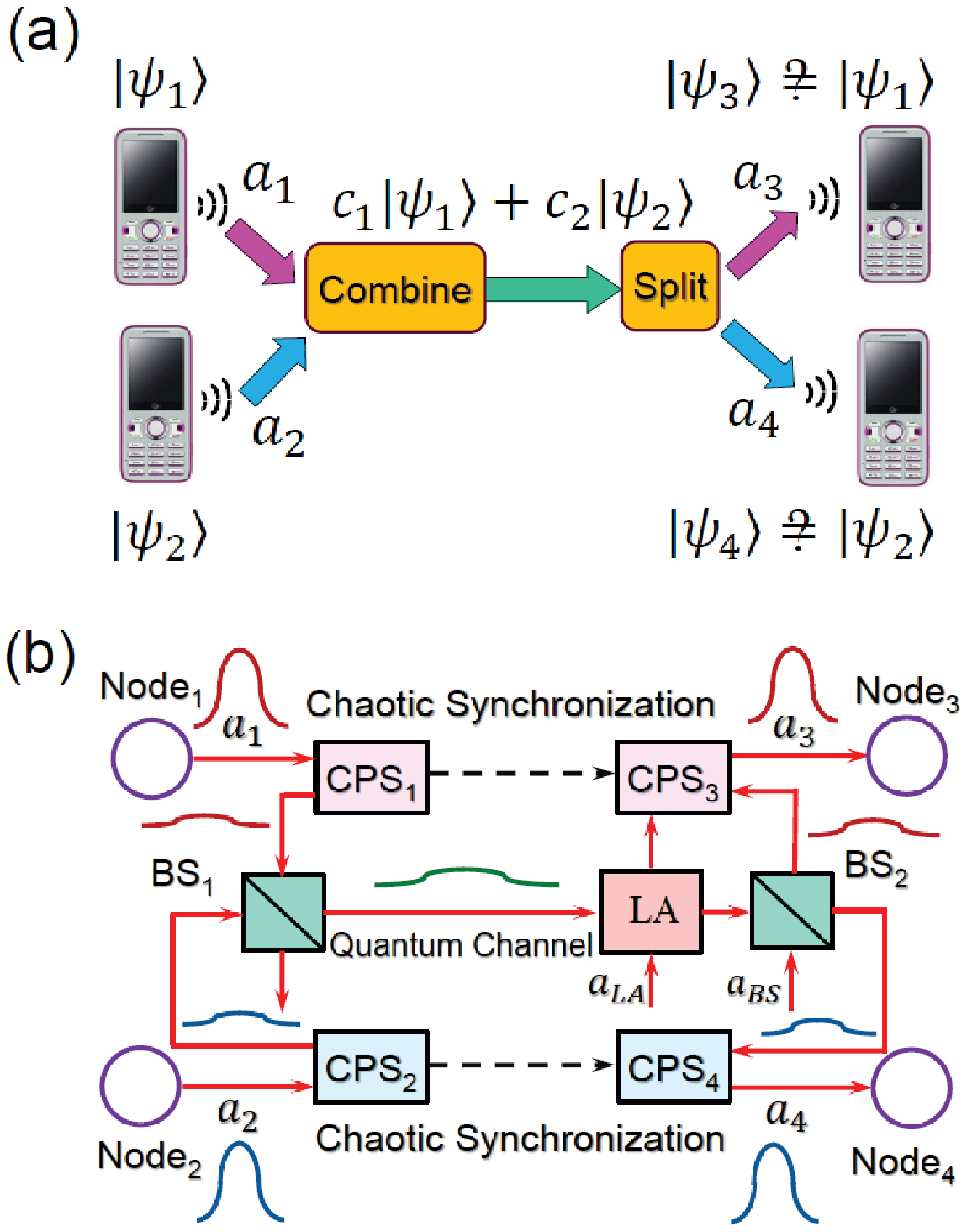,width=12cm}
\end{center}\caption{}
\label{Fig of the schematic diagram of quantum multiple access
network}
\end{figure}

\clearpage
\begin{figure}
\begin{center}
\epsfig{file=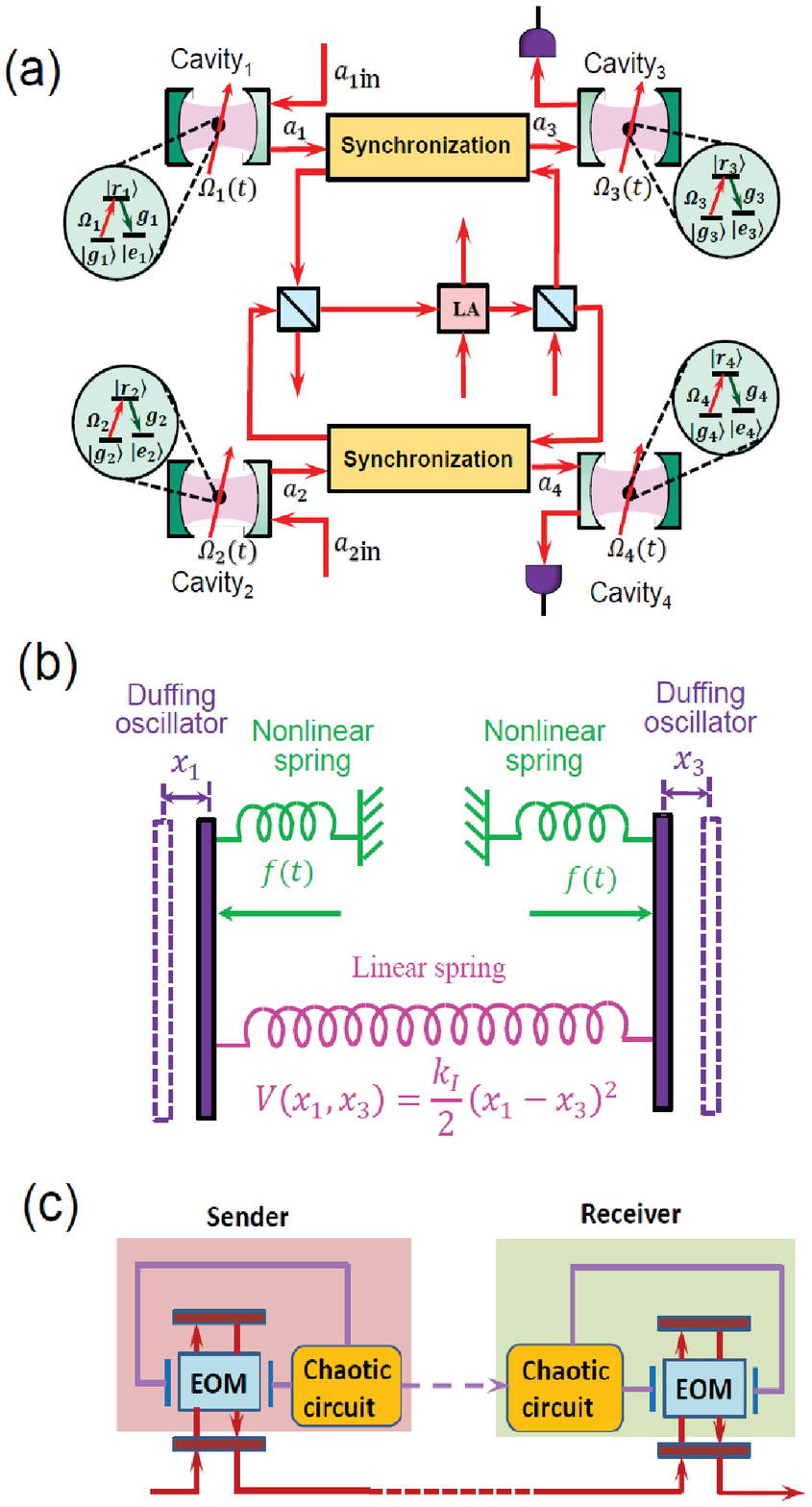,width=10cm}
\end{center}\caption{}
\label{Schematic diagram of quantum state transmission over
quantum CDMA network}
\end{figure}
\clearpage

\begin{figure}
\begin{center}
\epsfig{file=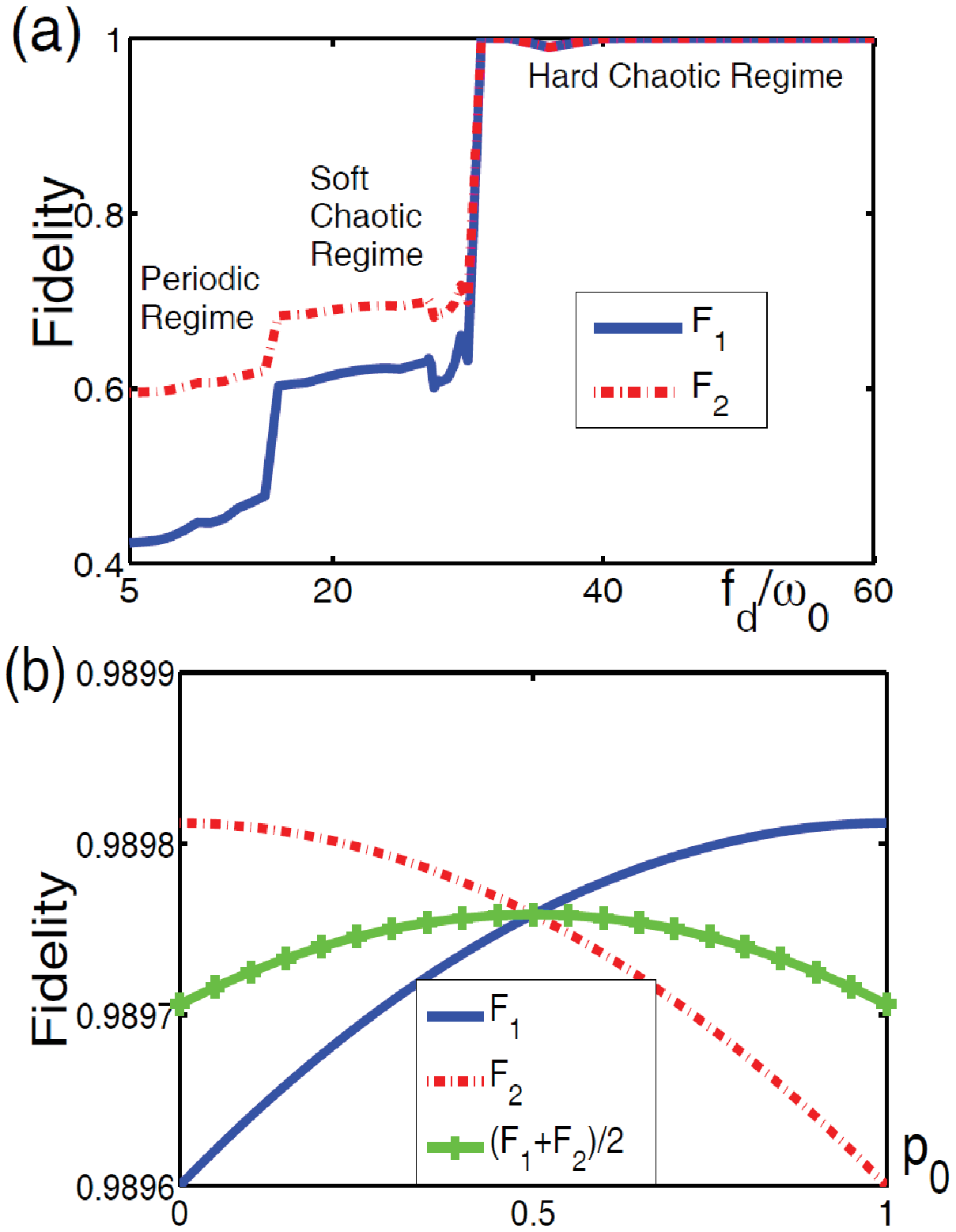,width=12cm}
\end{center}\caption{}
\label{Fig of the numerical results}
\end{figure}
\clearpage

\begin{figure}
\begin{center}
\epsfig{file=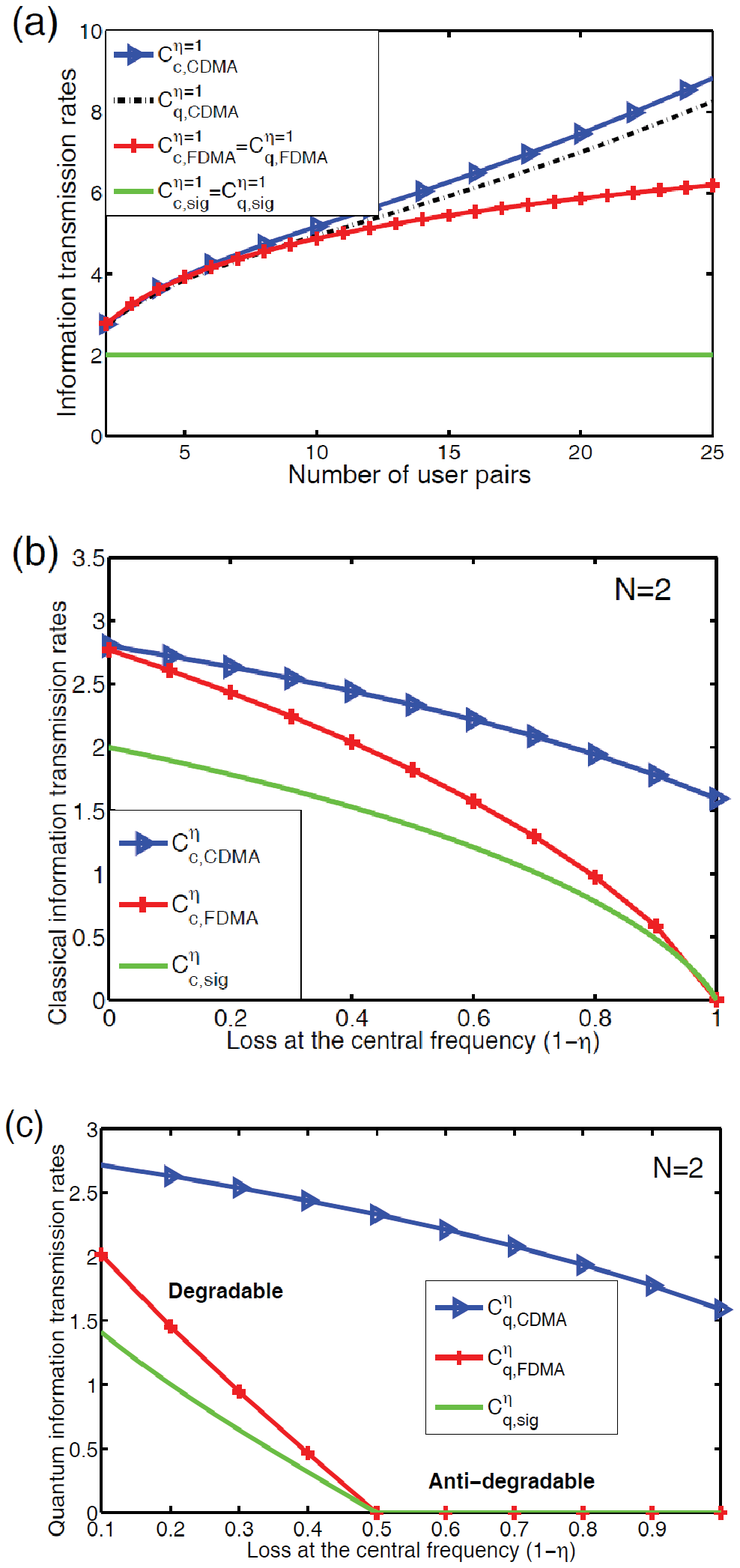,width=10cm}
\end{center}\caption{}
\label{Fig of the numerical results for information transmission
rates}
\end{figure}
\clearpage

\begin{figure}
\begin{center}
\epsfig{file=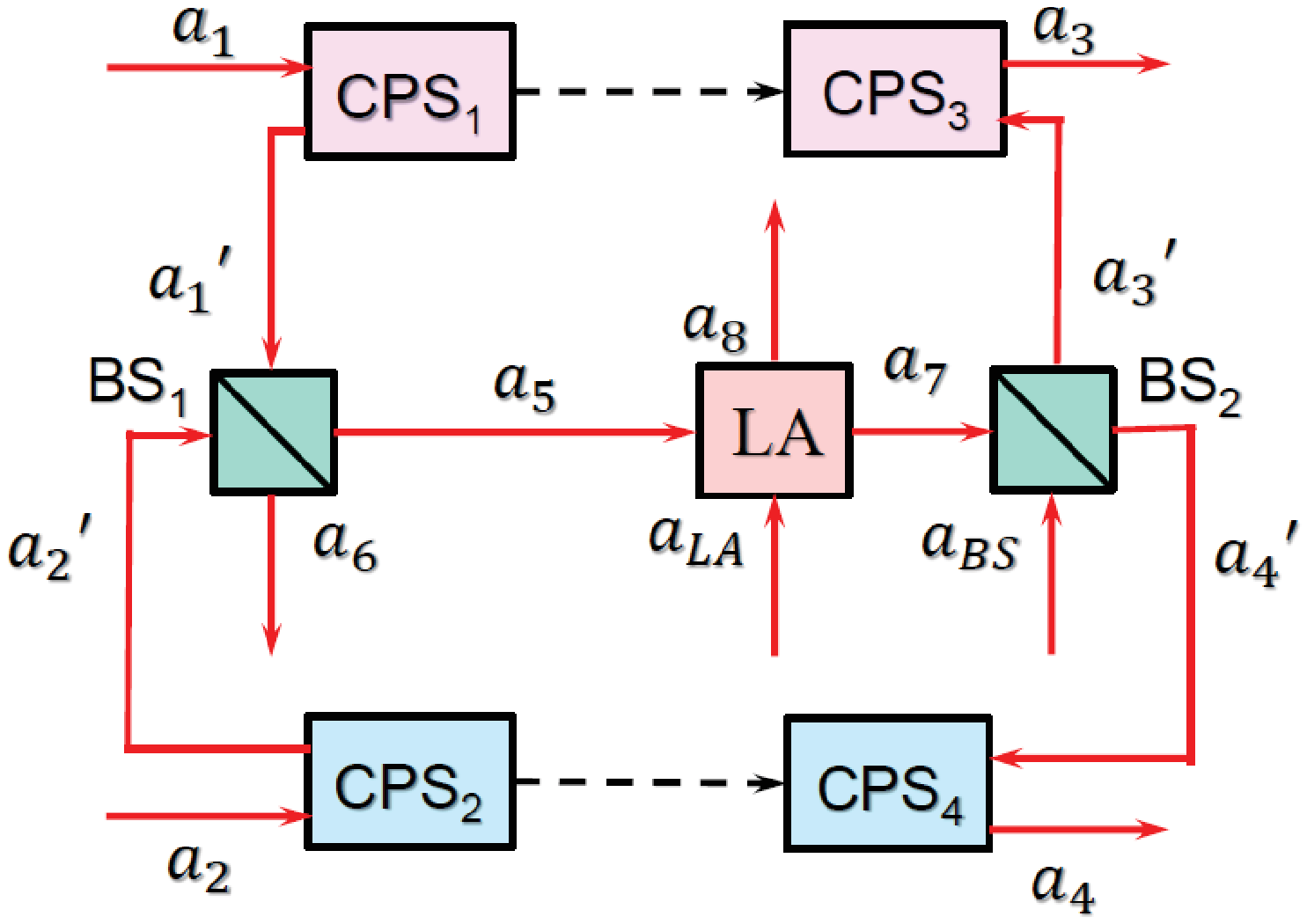,width=12cm}
\end{center}\caption{}
\label{Schematic diagram of quantum multiple access network with
chaotic phase shifters}
\end{figure}

\end{document}